# Network Structures between Strategies in Iterated Prisoners' Dilemma Games

Young Jin Kim, Myungkyoon Roh, and Seung-Woo Son

*Department of Applied Physics, Hanyang University, Ansan 426-791, Korea*

We use replicator dynamics to study an iterated prisoners' dilemma game with memory. In this study, we investigate the characteristics of all 32 possible strategies with a single-step memory by observing the results when each strategy encounters another one. Based on these results, we define similarity measures between the 32 strategies and perform a network analysis of the relationship between the strategies by constructing a strategies network. Interestingly, we find that a win-lose circulation, like rock-paper-scissors, exists between strategies and that the circulation results from one unusual strategy.

PACS numbers: 02.50.Le, 87.23.Ge, 87.23.Kg

Keywords: Prisoner's dilemma game, Complex networks, Strategies

Email: sonswoo@hanyang.ac.kr

Fax: +82-31-400-5473

## I. INTRODUCTION

Recently, a great deal interdisciplinary research has been performed crossing many different fields. Specifically, various methods developed in statistical physics have been introduced to understand the underlying mechanisms that successfully describe social, biological, and ecological phenomena, such as evolutionary biology [1, 2], agent-based models [3], voter models [4], and epidemic spreading [5]. Game theory is one of the prevalent topics in complexity studies, and typical examples are prisoner's dilemma, chicken games, public goods games, snowdrift games, and hawk-dove games [6-8].

Among the various examples, the prisoner's dilemma (PD) is the best known model [7, 8]. Two criminals are arrested and imprisoned. Because each prisoner is in solitary confinement, they have no way to deliver a message to or talk with each other. The police offer each prisoner a separate Faustian bargain. If one confesses and the other denies the crime, the confessor will be set free, and the other will serve three years in prison. If both confess, they will each serve two years in prison. On the other hand, if both deny the crime, each will serve only one year in prison. Since confession is always better than denial for each prisoner independently of whatever the partner's choice, both prisoners may confess. This is called the prisoner's dilemma because the two-year sentence for both is the worst case possibility for the local optimum of self-interested prisoners. The global optimum is both denying the crime, with each serving only one year.

Table 1. Payoff matrix for the player (opponent) in a PD game.

|  |  | opponent | |
|---|---|---|---|
|  |  | C | D |
| player | C | 3 ( 3 ) | 0 ( 5 ) |
|  | D | 5 ( 0 ) | 1 ( 1 ) |

However, if this game is repeated, what will happen? This is called the iterated prisoner's dilemma game (IPDG) [7]. The results show very different patterns from a single-iteration game. Sometimes, both prisoners deny the crime by making promises to each other in advance of getting caught. In some cases, one confesses and the other denies, like showing altruism depending on the details of rule. In this paper, we study the IPDG with a one-step memory size by using replicator dynamics [8, 9]. In section II, all 32 possible strategies are characterized and analyzed by applying a similarity measure. Based on the results of one-to-one games among the 32 strategies, we reveal the network structures between the strategies in Section III. Interestingly, a win-lose circulation exists between strategies. A summary of the results and discussions are addressed in the last section.

## II. METHODS

For simple notation, we denote denial with the capital letter 'C' meaning "cooperation", and confession with the letter 'D' meaning "defect", as conventionally used in game theory. If both players (prisoners) cooperate, they both receive a reward (R) for cooperation. If one defects while the other cooperates, then the defector receives a temptation (T) while the partner receives a sucker's payoff (S). If both players defect, both players receive punishment (P). For convenience, we set the payoffs at T=5, R=3, S=0, and P=1, as shown in Table 1, where the condition $T > R > P > S$ must hold for the game to be a PD game. Additionally, $2R > T + S$ is required to keep the mutual cooperation from alternating between cooperation and defection in the iterative game. Depending on the relation between the values of T, R, P, and S, the snowdrift game, the chicken game, and the hawk-dove game branch off [8].

## 1. Thirty-two Possible Strategies with a Memory Size One

The dependences of the characterized reaction patterns of players in the IPDG on former results (histories) are called "strategies." Because we consider the case of memory size one, i.e., only the previous step is remembered, we have five choices for each strategy. First, we have to consider the four possible previous results: both the player and the opponent cooperate in the previous step (denoted as |CC|), |CD| representing the case in which the player cooperates but the opponent defects, |DC|, and |DD|. Added to these, we should also consider the initial choice. Therefore, the number of all possible strategies is $2^5 = 32$ when taking two possible players' choices into account. If the convenient bitwise representation in Ref. 10 is borrowed, each strategy is coded by five bits, $a_0|a_1a_2a_3a_4$. The first bit, $a_0$, is the initial choice, and bit $a_1$ is the reaction for |CC|. Likewise, $a_2$ is for |CD|, $a_3$ for |DC|, and $a_4$ for |DD|. For example, the most famous strategy 'tit for tat' (TFT) is C|CDCD, which is the coding for the behavior in which the player first cooperates and then copies the opponent's choice [7, 8]. We assign a decimal number for each strategy by interpreting the binary notation 'C' as 1 and 'D' as 0 and by reading it in the reverse way. TFT corresponds to the 11[th] strategy because the binary number $01011_{(2)}$ is 11 in the decimal system. Other examples are 'grim trigger' (GT) and 'Pavlov.' The GT player cooperates first; then, if the opponent defects once, the player never forgives [8]. The bitwise representation C|CDDD describes GT, and corresponds to $00011_{(2)}$, i.e., the 3[rd] strategy. Pavlov strategy is 'win-stay-lose-shift,' whose bitwise

Table 2. Detailed behaviors of the 32 strategies.

| Strategy number | Bitwise representation | Strategy nickname | Behavior pattern description |
|---|---|---|---|
| 0 | D\|DDDD | All D | Always defect |
| 1 | C\|DDDD | All D FC | First cooperate; after then always defect |
| 2 | D\|CDDD | All D | Always defect |
| 3 | C\|CDDD | Grim trigger | First cooperate, then if the opponent defects once, player never forgives |
| 4 | D\|DCDD | All D | Always defect |
| 5 | C\|DCDD | | First cooperate; then only cooperate for \|CD\| |
| 6 | D\|CCDD | All D | Always defect |
| 7 | C\|CCDD | All C | Always cooperate |
| 8 | D\|DDCD | Swindler FD | Always defect except in a \|DC\| situation |
| 9 | C\|DDCD | Swindler | First cooperate; then always defect except in a \|DC\| situation |
| 10 | D\|CDCD | Tit for tat FD | First defect; then copy the opponent's choice |
| 11 | C\|CDCD | Tit for tat | First cooperate; then copy the opponent's choice |
| 12 | D\|DCCD | | First defect; if the opponent's choice is different, cooperate |
| 13 | C\|DCCD | | First cooperate; if the opponent's choice is different, cooperate |
| 14 | D\|CCCD | Blinder | First defect; if the opponent cooperates once, always cooperate |
| 15 | C\|CCCD | All C | Always cooperate |
| 16 | D\|DDDC | | First defect; only cooperate for \|DD\| |
| 17 | C\|DDDC | | First cooperate; only cooperate for \|DD\| |
| 18 | D\|CDDC | Pavlov FD | First defect; if the result is bad, flip the choice |
| 19 | C\|CDDC | Pavlov | First cooperate; if the result is bad, flip the choice |
| 20 | D\|DCDC | Tree frog FD | First defect; then choose choice opposite your opponent's |
| 21 | C\|DCDC | Tree frog | First cooperate; then choose choice opposite your opponent's |
| 22 | D\|CCDC | Prodigal son | First defect; if once punished, always cooperate |
| 23 | C\|CCDC | All C | Always cooperate |
| 24 | D\|DDCC | CD repeater | First cooperate; then change choice independently of the opponent's |
| 25 | C\|DDCC | DC repeater | First defect; then change choice independently of the opponent's |
| 26 | D\|CDCC | Punisher FD | First defect; defect only for the \|CD\| situation |
| 27 | C\|CDCC | Punisher | First cooperate; defect only for the \|CD\| situation |
| 28 | D\|DCCC | Rebel FD | First defect; defect only for the \|CC\| situation |
| 29 | C\|DCCC | Rebel | First cooperate; defect only for the \|CC\| situation |
| 30 | D\|CCCC | All C FD | First defect; then always cooperate |
| 31 | C\|CCCC | All C | Always cooperate |

representation is C|CDDC, the 19[th] strategy [9]. All 32 possible strategies are summarized in Table 2.

As one can see in Table 2, similarities exist among the 32 strategies. For example, 0, 2, 4, and 6 strategies are D|DDDD, D|CDDD, D|DCDD, and D|CCDD in the bitwise representation. Even though the bitwise representation (genotype) is different, all of these strategies only repeat 'D.' We call it 'All D' in their reaction (phenotype). Similarly, there are 4 genotypes for 'All C,' 7, 15, 23, and 31, and 2 genotypes for 'CD repeater,' 24, and 'DC repeater,' 25, which are effectively the same. We remove the seven effectively similar strategies later for analysis of the relation between the remaining 25 strategies.

In order to compare the characteristics of the 32 strategies, we calculate the cumulative points of each strategy against the others over 24 rounds based on the payoff matrix shown in Table 1, discarding the first 24 transient rounds. The resulting cumulative points are visualized in Fig. 1.

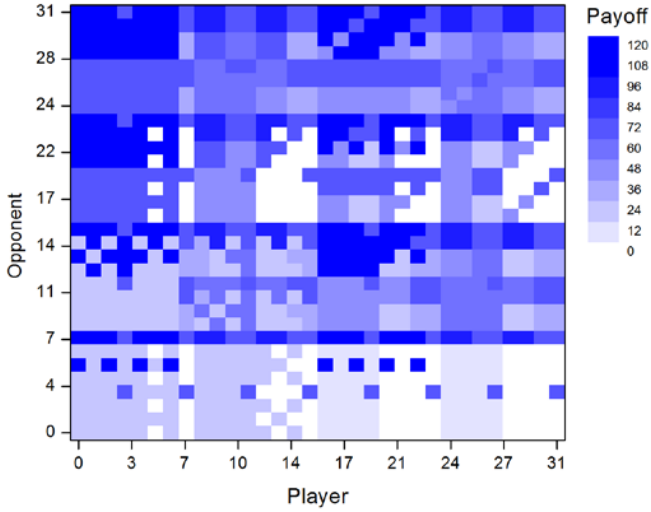

Fig. 1. (Color online) Cumulative payoff results for a one-to-one iterated game.

Clearly, one can see that effectively similar strategies exist. Using this cumulative point matrix, we measure the similarity between the 32 strategies by adopting the cosine inner product [11] and the Pearson correlation [12], which are the simplest ways to measure the correlation, to get a similarity between the strategies. The results using the cosine inner product provide results that are almost the same as that provided by using the Pearson correlation.

## 2. Replicator Dynamics

For the 25 non-duplicated strategies of the IPDG, we perform a replicator dynamics (RD) simulation, which is widely used to find evolutionarily stable strategies in evolutionary game dynamics [8, 9]. The population density of strategy $i$ is denoted by $x_i$, and its time derivative is given by the following equation:

$$\dot{x}_i = x_i[f_i(x) - \bar{f}(x)],$$
$$\bar{f}(x) = \sum_{j=1}^{n} x_j f_j(x), \quad (1)$$

where $f_i$ is the fitness (payoff) of strategy $i$ for a given population condition $x$. The above equation describes a relative growth rate of a strategy that is proportional to its relative payoff derived from the average over the whole population. In this simulation, we adopt a first-order Euler method to solve an ordinary differential equation with a time-change $dt = 0.01$. For each time step, in order to get the average fitness, we discard the transient payoffs over 24 rounds.

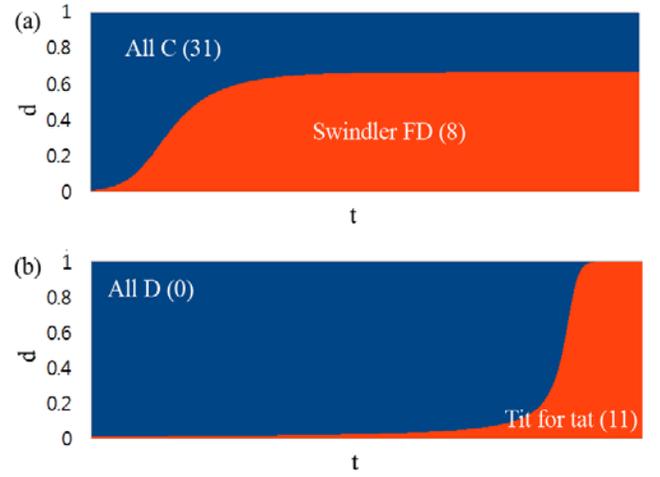

Fig. 2. (Color online) Typical examples of replicator dynamics simulations: (a) 1% of Swindler FD (8) invades 99% of All C (31) and then reaches the equilibrium point, and (b) 1% of tit for tat (11) invades 99% of All D and then overwhelms the majority.

For the 25 non-duplicated strategies, we perform extensive numerical simulations with two initial conditions. One is starting the 25 strategies in a homogeneous population and the other is one-to-one invading condition, 1% of the population of a certain strategy invades into a 99% defensive major strategy. Figure 2 shows the usual examples of a one-to-one invading game. In the case of replicator dynamics of the two strategies, one can easily solve the ordinary differential equations analytically. If we define fitness $F_I \equiv P_{II} - P_{DI}$ and $F_D \equiv P_{DD} - P_{ID}$, where $P_{ij}$ means the average payoff that strategy $i$ gets from the strategy $j$, Eq. (1) gives

$$\dot{x}_I = -(F_I + F_D)\left(x_I - \frac{F_D}{F_I + F_D}\right)x_I(x_I - 1), \quad (2)$$

where the subscripts I and D represent invading and defensive strategies, respectively. According to the signs and the amplitudes of $F_I$ and $F_D$, Eq. (2) has four significant scenarios: (a) stable extinction, (b) initial-condition-dependent an unstable point, (c) balanced equilibrium, and (d) overwhelming conquest, as shown in Fig. 3. We check that our numerical simulations agree with theoretical expectations.

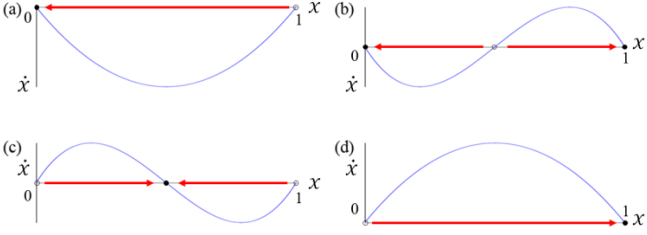

Fig. 3. (Color online) Four possible cases of replicator dynamics for one-to-one invading games: (a) stable extinction, (b) bistable extinction depending on the initial condition, (c) balanced equilibrium, (d) overwhelming dominant cases.

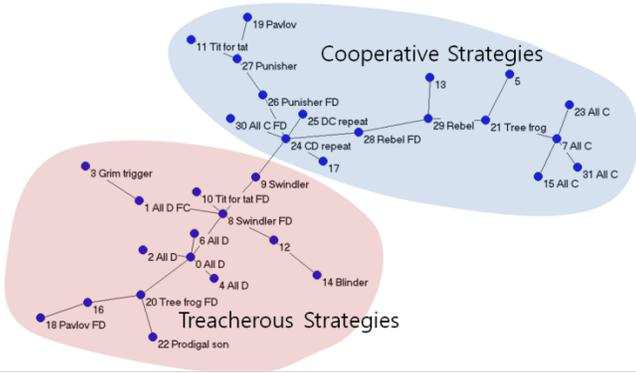

Fig. 4. (Color online) Maximum spanning tree graph based on the similarity measure.

## III. RESULTS

### 1. Strategy Similarity Map Using a Maximum Spanning Tree

Based on the cosine similarity measure of the matrix in Fig. 1, we construct a maximum spanning tree graph incorporating Prim's algorithm [13], as shown in Fig. 4. In this figure, cooperative strategies are well separated from treacherous strategies. The boundary strategy is the 'CD repeater,' number 24, which is the "neutral strategy" ignoring the opponent's choice. The effectively similar strategies mentioned above are clustered well in the tree graph. One may wonder why GT is classified as treacherous in Fig. 5. Even though GT cooperates first and continues the cooperation before the opponent defects, contrary to TFT, GT never forgives if the opponent defects once. Due to the consecutive defects, GT is classified as a treacherous strategy like the strategy all D except for the first cooperation.

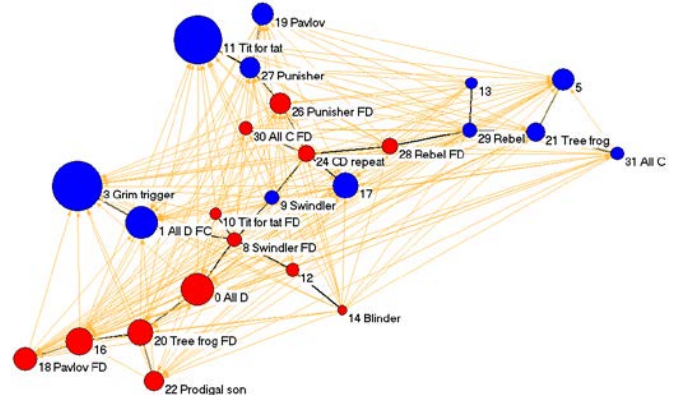

Fig. 5. (Color online) Strategy network. Blue and red nodes mean first moves are cooperative and defeatist respectively, black edges are the maximum spanning tree based on similarity measure, and orange directed edges are results of one-to-one replicator dynamics. The size of the node is proportional to $\frac{25}{(1+L)} + \frac{W}{3}$, where L is the number of losses and W is the number of wins.

### 2. Numerical Simulation for a Homogeneous Population

Putting the 25 non-duplicated strategies uniformly in the RD system, we run the simulation. The numerical results smoothly reach an equilibrium state without any rough oscillation. The most successful strategies are GT (number 3, 65.5%), TFT (number 11, 14.9%), Pavlov (number 19, 11.5%), Punisher (number 27, 7.5%), and All C (number 31, 0.6%). These results are similar to them reported in Ref. 10, even though the final population densities are a little different. The other strategies are less than 1% or extinct. Because the seven removed strategies are All C, All D, and CD repeater, the simulation results are not much different from there in Ref. 10.

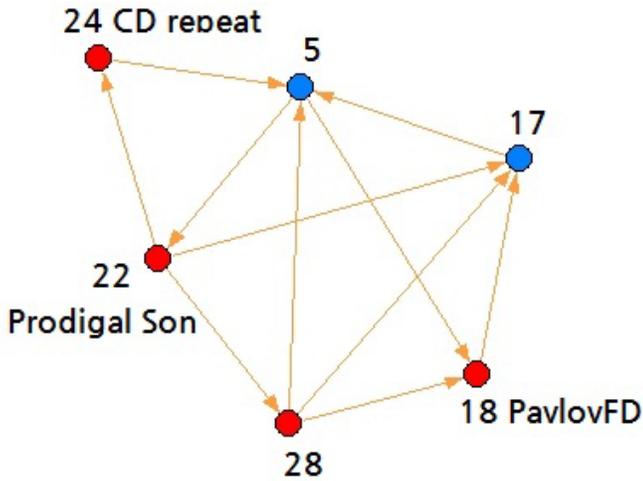

Fig. 6. (Color online) Win-lose circulation.

3. **Win-lose Network Structure between 25 Strategies**

From the theoretical calculation and numerical simulation of replicator dynamics between two strategies, we express all win-lose relations between two strategies among the 25 strategies, as shown in Fig. 5. The heads of the orange arrows indicate that one strategy always overwhelmingly dominates the other strategy. The black back-bone is the similarity map in Fig. 4. The color of the nodes represents the first move of the strategy, 'C' in blue and 'D' in red. The sizes of the nodes denote the estimated powers of the strategies based on the number of wins and losses. Similar to the above result, the TFT and the GT still have a strong influence, even in one-to-one games.

From these entangled arrows, we find an unexpected win-lose circulation, like rock-paper-scissors. Figure 6 shows the possible win-lose circulation taken from the win-lose network structure. Interestingly, this win-lose circulation is sustained by the node of the strategy 5, C|DCDD, which is a strategy we cannot easily characterize. This strategy only cooperates when its previous choice is 'C' and the opponent's is 'D'; otherwise, it is always defection. If we remove strategy 5, all the win-lose circulation loops collapse. This win-lose circulation can make the system oscillate.

## IV. SUMMARY AND CONCLUSIONS

In this study, we investigate the relation between 32 strategies of an iterated prisoners' dilemma game with a memory size 1. From the strategy similarity map, we find that there are seven effectively similar strategies. After eliminating these 7 strategies by using the results of one-to-one iterated games, we construct a win-lose network structure between the 25 strategies, and find interesting win-lose circulation loops. Strategy 5, C|DCDD, is found to play a key role in this win-lose circulation. Because win-lose circulation loops, like rock-paper-scissors, exist, one can expect oscillating population densities under certain condition. However, when the simulation starts from a homogeneous population density, we cannot find any oscillating behavior in the population density because strategy 5 quickly decays. If one chooses the three strategies that make a cyclic triangle from Fig. 6, one can observe the oscillation of the population density for the proper initial condition. We suggest a further analysis of these oscillating behaviors for a future work.

## ACKNOWLEDGEMENTS

We would like to thank Seung Ki Baek and Beom Jun Kim for helpful discussions. This work was supported by a National Research Foundation of Korea (NRF) grant funded by the Ministry of Science, ICT & Future Planning (No. 2012R1A1A1012150).

# REFERENCES

[1] M. Ruse, *Monad to Man: The Concept of Progress in Evolutionary Biology* (Harvard University Press, Harvard, 2009).
[2] J. L. Hendrikse, T. E. Parsons, and B. Hallgrímsson, Evolution & Development **9**, 393 (2007).
[3] R. Axelrod, *The Complexity of Cooperation: Agent-based Models of Competition and Collaboration* (Princeton University Press, Princeton, 1997).
[4] K. Suchecki, V. M. Eguíluz, and M. San Miguel, Phys. Rev. E **72**, 36132 (2005).
[5] L. Giuggioli, S. Pérez-Becker, and D. P. Sanders, Phys. Rev. Lett. **110**, 58103 (2013).
[6] M. Doebeli and C. Hauert, Ecol. Lett. **8**, 748 (2005).
[7] R. Axelrod and W. D. Hamilton, Science **211**, 1390 (1981).
[8] G. Szabo and G. Fath, Phys. Rep. **446** (2007).
[9] J. Hofbauer and K. Sigmund, *Evolutionary Games and Population Dynamics* (Cambridge University Press, Cambridge, 1998).
[10] S. K. Baek and B. J. Kim, Phys. Rev. E **78**, 011125 (2008).
[11] A. Singhal, *Modern Information Retrieval: A Brief Overview*, IEEE Data Eng. Bull. **24**, 35(2001).
[12] J. L. Rodgers and W. A. Nicewander, Amer. Stat. **42,** 59 (1988)
[13] R. C. Prim, *Shortest connection networks and some generalizations*, Bell Syst. Tech. J. **36**, 1389 (1957).